\documentclass[]{tMPH2e}

\usepackage{amsmath}
\usepackage{amssymb}
\usepackage{epsfig}
\usepackage{multirow}
\usepackage{graphicx}

\newcommand {\dr}{{\mathrm d}\mathbf{r}}

\newcommand {\dd}{{\mathrm d}}
\newcommand {\rr}{\mathbf{r}}

\begin{document}

\markboth{Andrew J.\ Archer and Alexandr Malijevsk\'y}{Molecular Physics}

\articletype{RESEARCH ARTICLE}

\title{On the interplay between sedimentation and phase separation phenomena in two-dimensional colloidal fluids}

\author{Andrew J.\ Archer$^{a}$$^{\ast}$\thanks{$^\ast$Email: A.J.Archer@lboro.ac.uk
\vspace{6pt}} and Alexandr Malijevsk\'y$^{b,c}$\\\vspace{6pt}  $^{a}${\em{Department of Mathematical Sciences, Loughborough University,\\ Loughborough LE11 3TU, UK}};\\
$^{b}${\em{E. H\'ala Laboratory of Thermodynamics, Institute of
Chemical Process Fundamentals of ASCR, 16502 Prague 6, Czech
Republic}}; \\
$^{c}${\em{Department of Physical Chemistry, Institute of
Chemical Technology, Prague,\\ 166 28 Praha 6, Czech Republic }}   }

\maketitle

\begin{abstract}
Colloidal particles that are confined to an interface effectively
form a two-dimensional fluid. We examine the dynamics of such
colloids when they are subject to a constant external force, which
drives them in a particular direction over the surface. Such a
situation occurs, for example, for colloidal particles that have
settled to the bottom of their container, when the container is
tilted at an angle, so that they `sediment' to the lower edge of
the surface. We focus in particular on the case when there are
attractive forces between the colloids which causes them to phase
separate into regions of high density and low density and we study
the influence of this phase separation on the sedimentation
process. We model the colloids as Brownian particles and use both
Brownian dynamics computer simulations and dynamical density
functional theory (DDFT) to obtain the time evolution of the
ensemble average one-body density profiles of the colloids. We
consider situations where the external potential varies only in
one direction so that the ensemble average density profiles vary
only in this direction. We solve the DDFT in one-dimension, by
assuming that the density profile only varies in one direction.
However, we also solve the DDFT in two-dimensions, allowing the
fluid density profile to vary in both the $x$- and $y$-directions.
We find that in certain situations the two-dimensional DDFT is clearly
superior to its one-dimensional counterpart when compared with
the simulations and we discuss this issue.

\bigskip

\end{abstract}

\section{Introduction}

Whilst colloidal fluids are, of course, of interest because of
their importance to industry and every-day-life -- for example the
blood in our arteries and veins is a complex colloidal suspension --
well controlled model colloidal suspension are also studied in
order to further our fundamental understanding of fluid behaviour
in general. In contrast to atoms, colloids are often large enough
that they can be observed under the microscope and so much
important knowledge has been gleaned concerning the microscopic
properties of liquids in general from studying colloidal
suspensions. In fact, it has been found that much from the theory of simple
liquids \cite{Hansen2006tsl, BarratHansen2003} can be utilised for their description.

In recent years, the non-equilibrium properties of (colloidal) fluids
on microscopic length and time scales has attracted considerable
interest
\cite{Hansen2006tsl, dhont1996idc}. Dynamical density functional theory (DDFT)
\cite{marconi1999uat, marconi2000ddf, archer2004ddf,
archer2004ddf_b} presents a particularly promising approach for
such study. DDFT provides a theory for the time evolution of Brownian
particles with Langevin stochastic equations of motion and therefore
colloidal models serve
as the natural subject for the theory. Recent work in this
direction includes studies of colloidal sedimentation such as that
described in Ref.\ \cite{royall2007nsc}, in which DDFT was found
to accurately describe the time evolution of the density profile
of sedimenting colloids measured using confocal microscopy, down
to the scale of the individual colloids. In this system, the
effective interactions between the colloids are purely repulsive
and well-modelled by the hard-sphere pair-potential and therefore
the sole mechanism for the aggregation of the colloids at the
bottom of the sample is the gravity driven sedimentation of the
colloids. However, in suspensions where the effective interactions
between the colloids includes attraction between the particles,
the aggregation can also be driven by these attractions. Such
attraction can be generated by the addition of non-adsorbing
polymer to the solution, which results in attraction due to the
so-called depletion mechanism
\cite{Poon2002jpcm, Rothetal2000pre}, and can lead to
the suspension phase separating into a colloid rich `liquid' phase
and a colloid poor `gas' phase. Such a system has, for example,
been used to study how droplets of the colloidal liquid phase
coalesce with the bulk of this phase that has already sedimented
to the bottom of the container \cite{AartsLekk2008jfm}. The benefit of studying
this phenomenon in a colloidal fluid is that it occurs on length
and time scales accessible to optical microscopy. The authors of
Ref.\ \cite{AartsLekk2008jfm} were able to study in detail the
various stages of the drop coalescing with the bulk fluid.

There is also a great deal of interest in two-dimensional (2D)
colloidal fluids. Such systems occur when the colloids become
confined to an interface. This can be at a fluid interface, such
as the air-water interface, where the colloids often become pinned
to the interface by capillary forces \cite{BresmeOettel2007jpcm,
Archer28}. Alternatively, a 2D system may be obtained when the
colloids settle under gravity to the bottom of their container.
Once on the bottom, the colloids are still mobile and able to
diffuse over the surface, but because the height above the
interface $h_{th}$ that they are able to reach due to the thermal
fluctuations is much less than the colloid diameter $\sigma$ at
common temperatures, they can effectively be considered to be a 2D
fluid of particles constrained to move on the bottom surface of
the container. The height $h_{th}\sim k_BT/m_Bg$, where $k_B$ is
Boltzmann's constant, $T$ is the temperature, $m_B$ is the buoyant
mass of the colloids and $g$ is the acceleration due to gravity.
In Refs.\ \cite{KoppletalPRL06, HenseleretalPRE10} the authors
studied the flow behaviour of such a 2D fluid down a narrow
channel with a width of only a few times the diameter of the
particles. In this set up, the colloids are ordered into layers
within the channel and a reduction in the number of layers due to
a density gradient along the channel was observed.


\begin{figure}
\centering
\includegraphics[width=0.4\columnwidth]{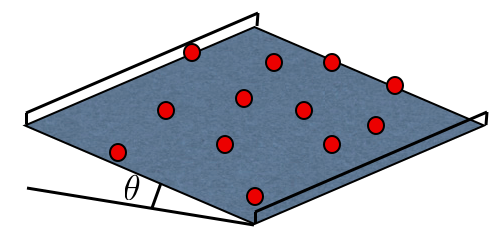}
\includegraphics[width=0.4\columnwidth]{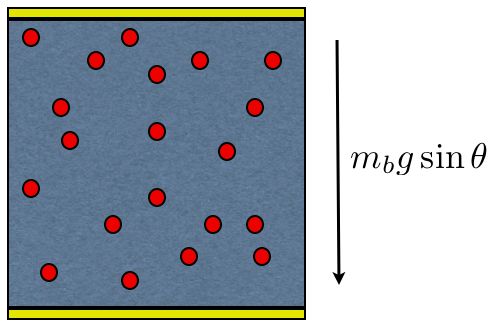}
\caption{The model system that we consider, consisting of colloidal particles
distributed over a surface. On the left we display a side view of
the set-up and on the right we display the view from above. When
the surface is tilted  at an angle $\theta$ to the horizontal, the
colloids experience a force $m_Bg\sin \theta$ parallel to the surface,
which causes them to `sediment' down to the bottom edge of the
surface.} \label{fig:set-up}
\end{figure}

In this paper, we present results obtained for a model 2D system
that is related to the system studied in Refs.\
\cite{KoppletalPRL06, HenseleretalPRE10}. In Fig.\
\ref{fig:set-up} we display a sketch of our system. We consider
cases where the colloids are initially uniformly dispersed over
the surface, with a fairly low average surface coverage. However,
due to the fact that there is a uniform external force on the
colloids towards one side of the surface, they begin to
`sediment'. In addition to this, there are attractive forces
between the colloids in our system. As the colloids move over the
surface, both due to the driving and also due to their Brownian
motion, they come into contact with one another and due to the
attractive interactions between the particles, they aggregate
together. We study the interplay between these two effects on the
time evolution of the density profile of the colloids using both Brownian
Dynamics (BD) computer simulations and DDFT.
We solve the DDFT in both one-dimension (i.e.\  assuming that the
density profiles only varies in the $y$-direction, which is the only direction
in which the external potential varies) and we also solve the DDFT in
2D, allowing the fluid density profiles to vary in both
the $x$- and $y$-directions. Perhaps the most striking result in this paper
is that we find that in certain situations the results from
solving the DDFT in 2D and then subsequently averaging over
the $x$-direction agree much better than the 1D DDFT results
with what we find from the BD simulations.

This paper is structured out as follows: In Sec.\ \ref{sec:mod_syst} we describe
our model system and give the governing stochastic equations of
motion for the particles. In Sec.\ \ref{sec:BD_res} we present BD simulation results
showing the typical behaviour of the system. Then, in Sec.\ \ref{sec:theory}
we present our theory for the system and give a brief discussion
of the equilibrium properties predicted by the theory.
In Sec.\ \ref{sec:results} we present our DDFT results and make
comparison with our BD simulation results. Finally, in Sec.\ \ref{sec:conc}
we make a number of concluding remarks.

\section{Model System}
\label{sec:mod_syst}

We model the colloids by assuming that they interact via the following pair potential:
\begin{equation}
v(r)=v_{hd}(r)+v_{at}(r),
\label{eq:pair_pot}
\end{equation}
where $r$ is the distance between the centres of the particles. The hard-disk pair potential is given by
\begin{equation}
v_{hd}(r) =
\begin{cases}
\infty \hspace{5mm} r \leq \sigma \\
0 \hspace{7mm} r > \sigma,
\end{cases}
\label{eq:pair_pot_hd}
\end{equation}
and the attractive contribution to the pair-potential is modelled by the following simple form:
\begin{equation}
v_{at}(r)=-\epsilon \exp(-r/l),
\label{eq:pair_pot_at}
\end{equation}
where $\epsilon$ is a parameter that determines the strength of
the attraction and $l$ sets the range of the potential. For
simplicity we set $l=\sigma$, the diameter of the hard-core of the
particles. Note that at contact $v(r=\sigma+)=-\epsilon/e$.

We model the external potential acting on the colloids as follows:
\begin{equation}
u_{ext}(x,y) =
\begin{cases}
\infty \hspace{5mm} y \leq 0 \\
ay \hspace{5mm} 0<y<L,\\
\infty \hspace{5mm} y \geq L \\
\end{cases}
\label{eq:ext_pot}
\end{equation}
where $y$ is the coordinate direction perpendicular to the hard
parallel walls that are located at $y=0$ and $y=L$ along the edge
of the surface on which the particles are located. The $y$-axis
is also parallel to the uniform external potential $ay$, where
$a=m_Bg\sin\theta$, for the set-up pictured in Fig.\ \ref{fig:set-up}.

We consider a system of $N$ particles and we denote the set of position
coordinates of the particles $\rr^N=\{\rr_i;\, i=1,\ldots,N \}$.
The potential energy of the system is given by
\begin{eqnarray}
U_N(\rr^N,t)
=\sum_{i=1}^N u_{ext}(\rr_i,t)
+ \frac{1}{2}  \sum_{i=1}^N \sum_{j \neq i} v(|\rr_i-\rr_j|).
\end{eqnarray}
We assume that the dynamics
of the colloids, with positions $\rr_i(t)$, is governed by the
following set of over-damped stochastic equations of motion
\cite{dhont1996idc}:
\begin{equation}
\Gamma^{-1}\frac{\dd \rr_i(t)}{\dd t}=-\nabla_i U_N(\rr^N,t)+ \xi_i(t),
\label{eq:EOM}
\end{equation}
where $\Gamma^{-1}$ is a friction coefficient characterising the
one-body drag of the solvent on the particles and $\xi_i(t)$ is a
stochastic white noise term \cite{dhont1996idc, marconi1999uat,
marconi2000ddf, archer2004ddf,
archer2004ddf_b}. Note that by modelling the equations of
motion via Eq.\ \eqref{eq:EOM} the hydrodynamic interactions
between the colloids are neglected \cite{dhont1996idc}.

\section{Brownian Dynamics Simulation Results}
\label{sec:BD_res}

\begin{figure*}
\centering
\includegraphics[width=0.4\columnwidth]{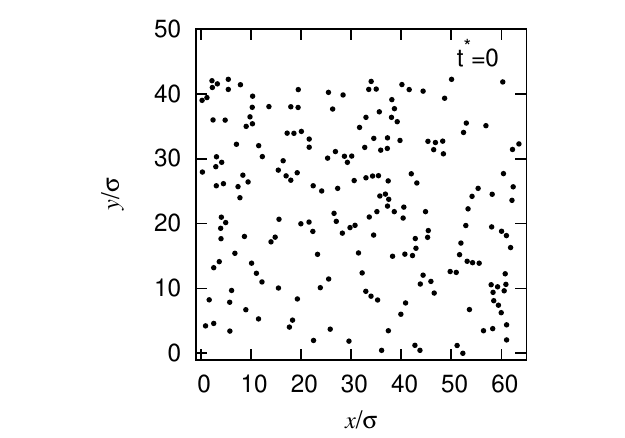}
\includegraphics[width=0.4\columnwidth]{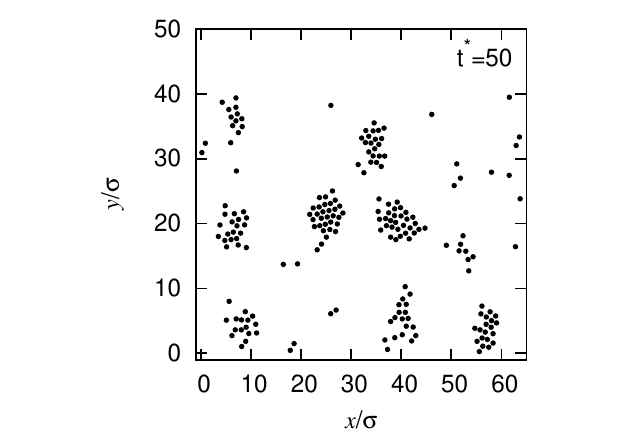}

\includegraphics[width=0.4\columnwidth]{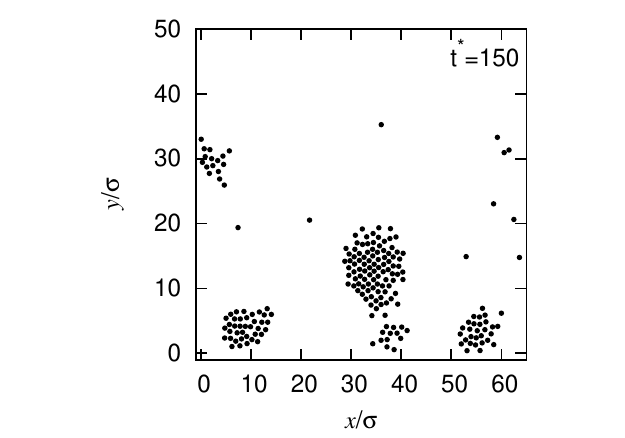}
\includegraphics[width=0.4\columnwidth]{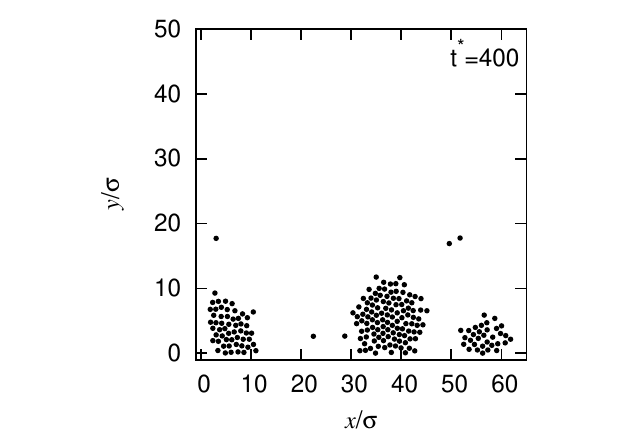}
\caption{\label{fig:snapshots}
Snapshots from our BD simulations
for the case when $L=43.5\sigma$, $\beta a=0.1$, $\beta \epsilon =
4.8$ and there are $N=200$ particles in the system, corresponding
to an average density $\bar{\rho}\sigma^2=0.072$. Results are
displayed at the times $t/\tau_B\equiv t^*=0$, 50, 150, and 400,
where $\tau_B=\beta \sigma^2/\Gamma$ is the Brownian
time-scale.}
\end{figure*}

\begin{figure*}
\centering
\includegraphics[width=0.4\columnwidth]{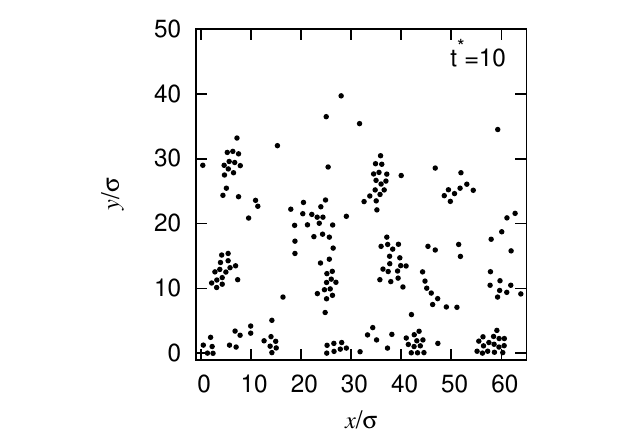}
\includegraphics[width=0.4\columnwidth]{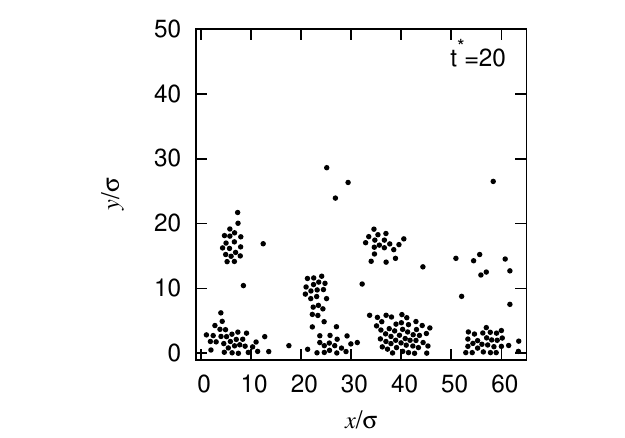}

\includegraphics[width=0.4\columnwidth]{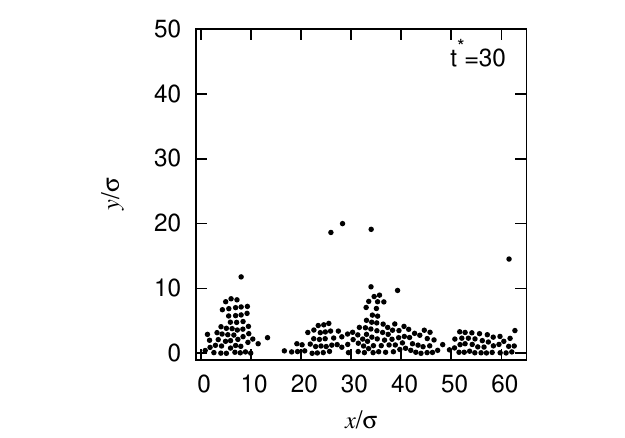}
\includegraphics[width=0.4\columnwidth]{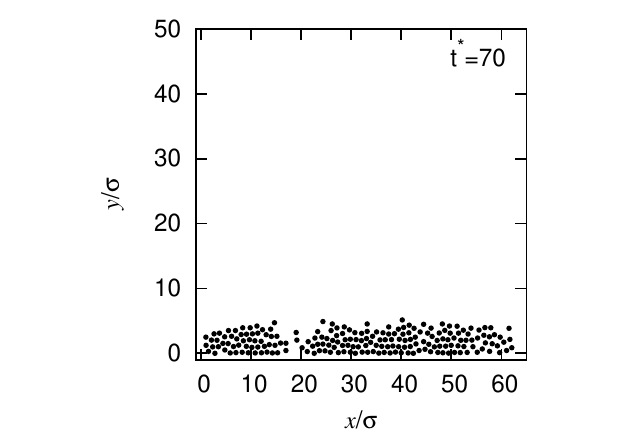}
\caption{\label{fig:snapshots2} Snapshots from our BD simulations
for the case when all the parameters (including the initial time $t=0$
starting positions of the particles)
are the same as in Fig.\ \ref{fig:snapshots}. the only difference is that
the driving force is one order of magnitude larger -- i.e.\
$\beta a=1$. Results are
displayed at the times $t/\tau_B\equiv t^*=10$, 20, 30, and 70,
where $\tau_B=\beta \sigma^2/\Gamma$ is the Brownian
time-scale.}
\end{figure*}

\begin{figure*}
\centering
\includegraphics[width=0.4\columnwidth]{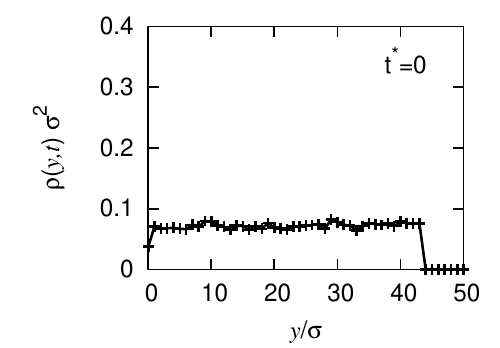}
\includegraphics[width=0.4\columnwidth]{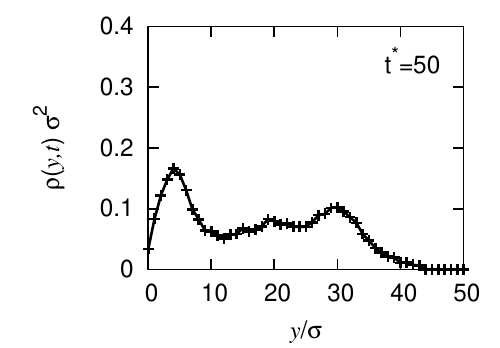}

\includegraphics[width=0.4\columnwidth]{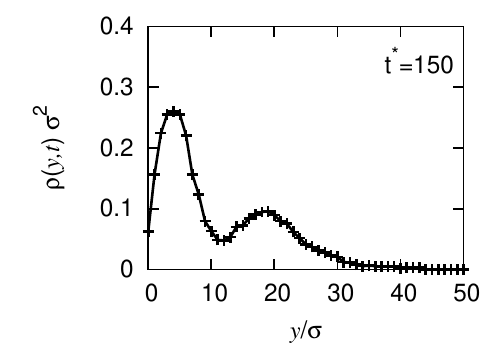}
\includegraphics[width=0.4\columnwidth]{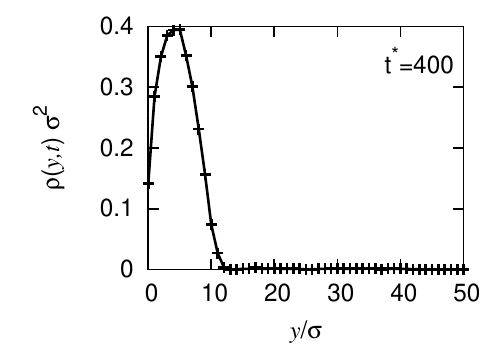}
\caption{\label{fig:BD_profiles} Time evolution of the fluid
density profiles $\rho(y,t)$, calculated from BD simulations, by averaging over
the $x$-direction. These results are obtained for the same
parameter values as the results displayed in Fig.\
\ref{fig:snapshots}. Results are displayed at the times
$t/\tau_B\equiv t^*=0$, 50, 150 and 400, where
$\tau_B=\beta \sigma^2/\Gamma$ is the Brownian time-scale.}
\end{figure*}

We have performed standard Brownian Dynamics computer simulations
\cite{AllenTildesley1989csl} of the time evolution of the particles in our system.
This corresponds to numerically integrating Eqs.\ \eqref{eq:EOM}.
To make this integration more straight-forward, we replace the
discontinuous hard-disk potential \eqref{eq:pair_pot_hd} with the
following continuous potential:
\begin{equation}
v_{hd}^*(r)=\alpha \exp(-(r/\sigma)^n).
\label{eq:pair_pot_hd_star}
\end{equation}
Choosing the amplitude $\beta\alpha=20$, where $\beta=1/k_BT$ is
the inverse temperature, together with a large value for the
exponent $n=50$ in $v_{hd}^*(r)$, ensures that using this
potential instead of the hard-disk-potential
\eqref{eq:pair_pot_hd} leads to no change in the structure of the
fluid.

We initiate our system with the non-overlapping particles randomly
distributed over the surface, which corresponds to an equilibrium
configuration for a system where there is no attraction between
the particles ($\epsilon=0$) and the drive force amplitude $a=0$,
corresponding to the particles being on a flat horizontal surface.

In Fig.\ \ref{fig:snapshots} we display snapshots
from a simulation for the case when $L=43.5\sigma$, $\beta a=0.1$,
$\beta \epsilon = 4.8$ and there are $N=200$ particles in the
system, corresponding to an average density
$\bar{\rho}\sigma^2=0.072$. Note that when $\beta \epsilon = 4.8$,
this corresponds to a potential well depth $\beta
v(r=\sigma+)=-\beta \epsilon/e=-1.77$. This amounts to a
relatively strong attraction between the particles and over time
it leads to the particles aggregating into clusters. We also see
in Fig.\ \ref{fig:snapshots} the influence of the external
potential which causes the particles to sediment to the bottom.
Due to the fact that there is no attraction between the particles
and the wall, the `drops' of the colloidal liquid phase do not
spread over the lower wall, so that it remains `dry'. From the
snapshots in Fig.\ \ref{fig:snapshots} we infer that in this
situation the contact angle is close to $180^o$. Note however,
that the contact angle decreases as the strength of the driving
force $a$ due to the external potential is increased. When $a$ is
increased to the value $\beta a=1$, then as can be observed from
the BD simulation results displayed in Fig.\ \ref{fig:snapshots2},
we find that the force on the particles towards the lower edge is
strong enough to cause them to spread to form a uniform layer
there. Thus, increasing the parameter $a$, which corresponds to
increasing the angle $\theta$ of the surface to the horizontal
(c.f.\ Fig.\ \ref{fig:set-up}), enhances the coalescence effect.

In Fig.\ \ref{fig:BD_profiles} we display the fluid one-body
density profiles $\rho(y)$, which are calculated by averaging the
full 2D density profile $\rho(x,y)$ over the $x$-direction. These
are obtained for the same parameters as for the results displayed
in Fig.\ \ref{fig:snapshots} by averaging over 40 different
simulation runs, each with a different initial configuration and
realisation of the random noise terms in Eq.\ \eqref{eq:EOM}. We
see that the contact density $\rho(y=0)$ at the lower wall, is
much less than the density just above the wall, at around $y\sim
3\sigma$. This low contact density comes as a consequence of the
fact that the amplitude $a$ of the driving force is small and also
that there are no attractive forces between the particles and the
wall and so the particles seek to gather to themselves away from the wall. The
surface tension (excess free energy) between the wall and the liquid is therefore very
similar in value to the liquid-gas surface tension between the
colloidal liquid and colloidal gas phases. As a consequence
of this and the fact that the surface tension between the vapour
and the wall is low, the drops on the surface displayed
in Fig.\ \ref{fig:snapshots} have a high contact angle, as one would expect from Young's equation.

\section{Theory for the system}
\label{sec:theory}

We use dynamical density functional theory \cite{marconi1999uat,
marconi2000ddf, archer2004ddf, archer2004ddf_b}, in order to
develop a theory to further understand the behaviour of our
system. Since DDFT is based on equilibrium density functional
theory (DFT) \cite{evans79, evans1992fif, Hansen2006tsl}, which is
a theory for the equilibrium one-body density profile $\rho(\rr)$
of an inhomogeneous fluid, we start this section by introducing
DFT and the approximation we use for the free energy functional.
Once set, DFT also allows one
to calculate thermodynamic quantities and phase behaviour for the
system.

\subsection{DFT}
\label{sec:DFT}

DFT shows that for a given external potential $u_{ext}(\rr)$,
there is a unique equilibrium one-body density profile $\rho(\rr)$ which is
obtained as the minimum of the following grand potential
functional \cite{evans79, evans1992fif, Hansen2006tsl}:
\begin{equation}
  \Omega[\rho(\rr)]=F[\rho(\rr)]-\mu\int \dr \rho(\rr),
\label{eq:Omega}
\end{equation}
where $\mu$ is the chemical potential and $F[\rho(\rr)]$ is the
Helmholtz free energy functional. Minimization leads to the
following Euler-Lagrange equation that may be solved for the
equilibrium fluid density profile:
\begin{equation}
\frac{\delta F[\rho(\rr)]}{\delta \rho(\rr)} = \mu.
\label{eq:F_min}
\end{equation}
The Helmholtz free energy functional for a 2D fluid is:
\begin{eqnarray}
  F[\rho(\rr)] = k_BT \int \dr \rho(\rr)
  [\ln(\Lambda^2\rho(\rr))-1] \notag \\
  +F_{ex}[\rho(\rr)]
  +\int \dr \, u_{ext}(\rr) \rho(\rr).
\label{eq:F_one}
\end{eqnarray}
The first term on the right hand side is the ideal-gas Helmholtz
free energy; $\Lambda$ is the thermal de Broglie wavelength.
$F_{ex}[\rho(\rr)]$ is the excess Helmholtz free energy, which
is the contribution to the free energy from the
particle-particle interactions. This quantity is not known
exactly, however, there are various well-trodden paths available
in the literature for obtaining a good approximation for this
quantity \cite{evans79, evans1992fif, Hansen2006tsl}. The final
term in Eq.\ \eqref{eq:F_one} is the contribution to the free
energy due to the external potential.

For the work described in this paper, we approximate the Helmholtz free energy functional as follows:
\begin{eqnarray}\notag
F[\rho(\rr)]&=&\int \dr f_{hd}(\rho(\rr))\\
&\,&+{ \frac{1}{2}\int \dr \int \dr' \rho(\rr)\rho(\rr')v_{at}(|\rr-\rr'|)} \notag \\
&\,&+\int \dr \rho(\rr)u_{ext}(\rr),
\label{eq:F_our}
\end{eqnarray}
where $f_{hd}(\rho)$ is the Helmholtz free energy per unit area of a uniform fluid of hard-disks with bulk density $\rho$. We use the scaled particle approximation \cite{rosenfeldPRA1990},
\begin{equation}
\beta f_{hd}(\rho)=\ln(\Lambda^2
\rho)-2-\ln(1-\eta)+(1-\eta)^{-1}, \label{eq:f_hd}
\end{equation}
which includes the exact ideal-gas contribution to the free energy
and where $\eta=\pi \rho \sigma^2/4$ is the packing fraction. The
second term in Eq.\ \eqref{eq:F_our} is a simple mean-field
approximation for the contribution to the free energy from the
attractive interactions between the particles. Note that the
approximation for $F[\rho]$ in Eq.\ \eqref{eq:F_our} does not give
reliable results for the fluid density profile when $\epsilon=0$,
or when the external potential due to the wall includes an
attractive tail. In this case there are oscillations in the
density profile near the wall, due to the effects from packing of
the hard cores of the particles against the wall. This is an
effect which Eq.\ \eqref{eq:F_our} is completely unable to
describe. However, for the present model fluid, the approximation
in Eq.\ \eqref{eq:F_our} is good, as long as $a$ is small ($\beta
a \sim 0.1$), and $\epsilon$ is sufficiently large ($\beta
\epsilon \gtrsim 1$), owing to the fact that as $\epsilon$ is
increased, the density of the particles in contact with the wall
is low, as discussed above in the context of the results displayed
in Figs.\ \ref{fig:snapshots} and \ref{fig:BD_profiles}.

\begin{figure}[t]
\centering
\includegraphics[width=0.5\columnwidth]{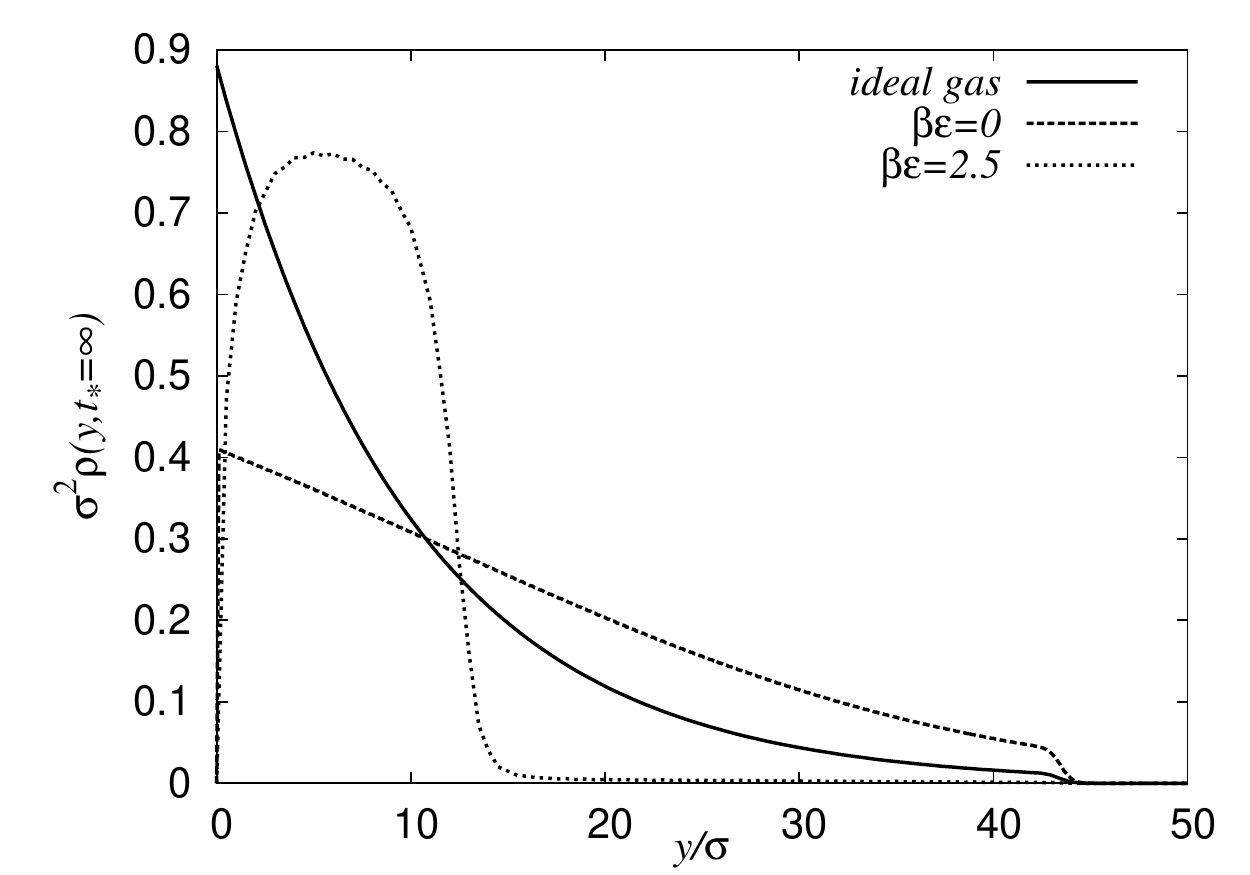}
\caption{\label{fig:comparison} Equilibrium ($t \to \infty$)
density profiles obtained using DFT for the case when the
parameters in the external potential are $L=43.5\sigma$, $\beta
a=0.1$ and when the average density in the system is
$\bar{\rho}\sigma^2=0.2$. We display the (exact) ideal-gas density
profile, together with the density profiles obtained using Eq.\ \eqref{eq:F_our}
when there is no attraction between the particles $\beta
\epsilon=0$ (pure hard-disks) and when the attraction amplitude is
$\beta \epsilon=2.5$.}
\end{figure}

To further illustrate this point, in Fig.\ \ref{fig:comparison} we display
the particle density profiles obtained using Eq.\ \eqref{eq:F_our} for the case when
the parameters in the external potential are $L=43.5\sigma$ and
$\beta a=0.1$ \footnote{Note that in our DFT and DDFT calculations we replace the upper hard wall at $y=L$ in Eq.\ \eqref{eq:ext_pot} by a softened purely repulsive form. This is done to improve the stability of our numerical algorithm for solving the DDFT. This, of course, has no effect on the behaviour of the system at the lower wall, which is the primary focus of our work described here.}. In our calculations we have set the value of the
chemical potential $\mu$ so that in all cases the average density
in the system is $\bar{\rho}\sigma^2=\frac{1}{L}\int_0^L dy \rho(y)
\sigma^2=0.2$. We display in Fig.\ \ref{fig:comparison} the
density profile for the case when $\beta \epsilon=2.5$, which
corresponds to a strong enough attraction between the particles
for the system to
exhibit liquid-gas phase separation and due to the external
driving force $a$ towards the bottom, we find the high density
liquid phase is pushed to the bottom of the system. However, the
density near to contact with the wall (at small values of $y$) is
low, which justifies our use of the local density approximation
(LDA) for the hard-disk contribution to the free energy in Eq.
\eqref{eq:F_our}. We also display in Fig.\ \ref{fig:comparison}
the (exact) ideal-gas density profile and the density profile
obtained using Eq.\ \eqref{eq:F_our} when there is no attraction between the
particles $\beta \epsilon=0$ (i.e.\ pure hard-disks). We see that
in this case the density profile near the wall is monotonic,
whereas in reality we should expect to see oscillations due to
packing near the wall \cite{evans1992fif}.

From Fig.\ \ref{fig:comparison} we can also observe the influence
of the nature of the particle interactions on the fluid density
profiles: on comparing the ideal gas density profile with the
$\epsilon=0$ pure hard-disk case, we see that the effect of the
excluded area (hard-core) interactions between the particles is to
push some of them further up into the system. However, the effect
of adding attraction between the particles is to reverse this
effect, so that when $\beta \epsilon=2.5$ the particles are on
average much closer to the bottom of the system.
This effect, of course, is also due to the external potential ($a>0$)
driving the particles to the bottom of the system.

\subsection{On the nature of the density profiles}
\label{sec:profile_nature}

There is an important issue that we must raise at this point in
our discussion: When a fluid is confined in an external potential
such as that given in Eq.\ \eqref{eq:ext_pot}, that only varies in one
direction (in this case the $y$-direction), then we know that the
fluid density profile $\rho(\rr)$ will only vary in this direction. This
is because $\rho(\rr)$ is an ensemble average quantity, which means that even
if particular configurations of the system are inhomogeneous in
the $x$-direction, such as in the configuration displayed in the final panel of
Fig.\ \ref{fig:snapshots} (which is an equilibrium configuration), the average over the full ensemble of
possible configurations will not vary in the $x$-direction. Thus,
in calculating the density profiles displayed in Fig.\
\ref{fig:comparison}, we have used this fact to reduce solving
Eq.\ \eqref{eq:F_min}, to that of solving for a one-dimensional (1D)
function $\rho(y)$.
However, although formally $\rho(\rr)$ does not vary in the
$x$-direction, as we show below, when one uses an {\em
approximate} free energy functional, such as that in Eq.\
\eqref{eq:F_our}, one can in fact obtain density profiles that
vary in both the $x$- and in the $y$-directions. We will further
discuss this issue below.

\subsection{Bulk phase diagram}
\label{sec:phase_diag}

\begin{figure}[t]
\centering
\includegraphics[width=0.5\columnwidth]{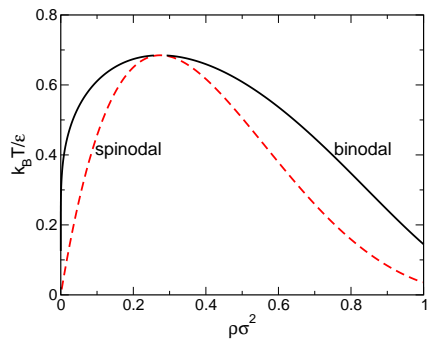}
\caption{\label{fig:phase_diag} Phase diagram in the reduced
temperature $k_BT/\epsilon$ versus density $\rho$ plane, obtained
from our simple approximation for the free energy in Eq.\
\eqref{eq:F_our}. Within the binodal curve the uniform fluid is
predicted to be unstable. Note that our simple theory is not
able to predict the existence of the hexatic or solid phases.}
\end{figure}

From our DFT \eqref{eq:F_our} we are able to calculate the bulk
(uniform) fluid phase diagram for our system, by setting the
density in Eq.\ \eqref{eq:F_our} to be a constant
$\rho(\rr)=\rho$, which gives the bulk fluid Helmholtz free
energy. From this the phase diagram may be calculated in the usual
way \cite{Chandlerbook}. The resulting phase diagram is displayed
in Fig.\ \ref{fig:phase_diag}. We see that as the temperature $T$
is decreased or, equivalently, as $\epsilon$ is increased, the
theory predicts that the system exhibits phase coexistence between
a low density suspension and a high density suspension of
particles (i.e.\ gas-liquid phase coexistence). The pairs of
coexisting state points define the locus of the binodal. The line
in the phase diagram along which the compressibility is predicted
to be zero defines the spinodal. Inside this curve the DDFT (see
Eq.\ \eqref{eq:DDFT} below) predicts that the uniform fluid is
linearly unstable. Note that our simple theory is not able to
predict the transition to either the hexatic phase nor the solid
phase, that we expect to find at higher densities. Note also that
the phase diagram in Fig.\ \ref{fig:phase_diag} is only
qualitatively correct in describing the gas-liquid phase
separation. For example, when $\beta \epsilon=4.8$, corresponding
to the results displayed in Figs.\ \ref{fig:snapshots} and
\ref{fig:BD_profiles}, we find that the density of the liquid at
coexistence is $\bar{\rho} \sigma^2\simeq 0.85$. However, from
Fig.\ \ref{fig:phase_diag} we see that we must select $\beta
\epsilon=3.5$ (i.e.\ $k_BT/\epsilon=0.29$) in order for the
density of the liquid at coexistence to equal this value. As a
consequence of this, we find below in Sec.\ \ref{sec:results}
better agreement between the DDFT results for $\beta \epsilon=3.5$
with the BD simulation results for $\beta \epsilon=4.8$, than we
do for the simulation results for $\beta \epsilon=3.5$.

\subsection{DDFT}
\label{sec:DDFT}

DDFT \cite{marconi1999uat, marconi2000ddf, archer2004ddf,
archer2004ddf_b} was developed to describe the dynamics of the
one-body density of a fluid of Brownian particles with equations
of motion given by Eq.\ \eqref{eq:EOM}. The Smoluchowski equation
\cite{dhont1996idc,archer2004ddf}:
\begin{eqnarray}
  \frac{\partial P^{(N)}}{\partial t} = \Gamma \sum_{i=1}^N
  \nabla_i \cdot [ k_BT \nabla_i P^{(N)}+ \nabla_i U_N P^{(N)} ]
  \label{eq:smoluchowski}
\end{eqnarray}
governs the time evolution of the $N$-particle probability density $P^{(N)}(\rr^N,t)$
for systems with dynamics governed by Eq.\ \eqref{eq:EOM}.
The one-body density is obtained by integrating over this function, as follows:
\begin{equation}
\rho(\rr_1,t) \, =\, N \int \dr_2 \, ... \int \dr_N P(\rr^N,t).
\end{equation}
This non-equilibrium density profile represents an ensemble average over
all realisations of the stochastic noise \cite{archer2004ddf_b}.
On integrating the Smoluchowski equation (\ref{eq:smoluchowski}) we
obtain the central equation of DDFT \cite{marconi1999uat,
marconi2000ddf, archer2004ddf, archer2004ddf_b}:
\begin{align}
  \frac{\partial \rho(\rr,t)}{\partial t} &= \Gamma \nabla \cdot
  \left[\rho(\rr,t) \nabla \frac{\delta F[\rho(\rr,t)]}{\delta
      \rho(\rr,t)}\right].
\label{eq:DDFT}
\end{align}
Note that this depends on the {\em equilibrium} Helmholtz free
energy functional $F[\rho]$, given in Eq.\ \eqref{eq:F_one}. In
order to derive Eq.\ \eqref{eq:DDFT}, we have made the {\em
approximation} that equal-time two-body correlations at time $t$
in the non-equilibrium fluid are the same as those of an
equilibrium fluid with the same one-body density profile
$\rho(\rr,t)$ \cite{marconi1999uat, marconi2000ddf,
archer2004ddf}. As we see below, this closure approximation is
good for the situation of interest here. Other situations where
DDFT (\ref{eq:DDFT}) has proved to be reliable are discussed in
Refs.\ \cite{marconi1999uat,
  dzubiella2003mfd, penna2003ddf, archer2005ddf, archer2005ddf_b,
  rex2005scd, rex2006ucc, royall2007nsc, rex2007ddf, rauscher2007ddf}. We should emphasise that in addition to the closure approximation
used in deriving the DDFT \eqref{eq:DDFT}, there is the additional
approximation that one must make in selecting a suitable
approximation for the Helmholtz free energy functional
$F[\rho(\rr)]$. The approximation that we use here is that in Eq.\
\eqref{eq:F_our}.

\begin{figure*}
\centering
\includegraphics[width=0.32\columnwidth]{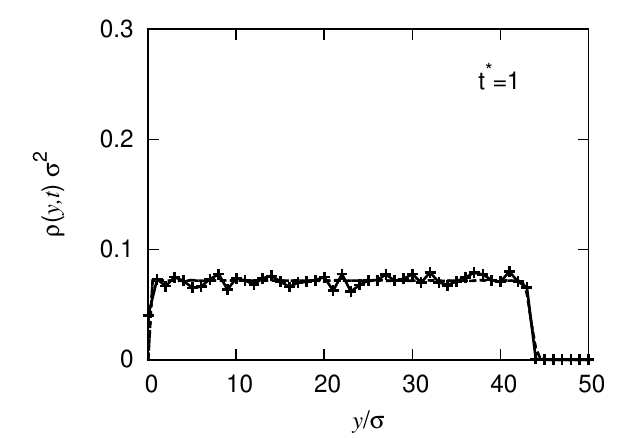}
\includegraphics[width=0.32\columnwidth]{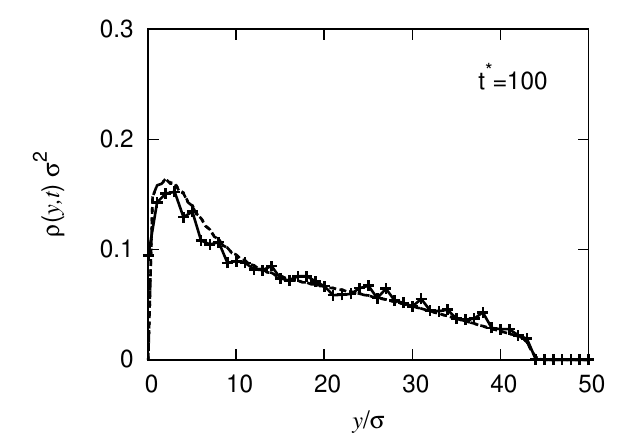}
\includegraphics[width=0.32\columnwidth]{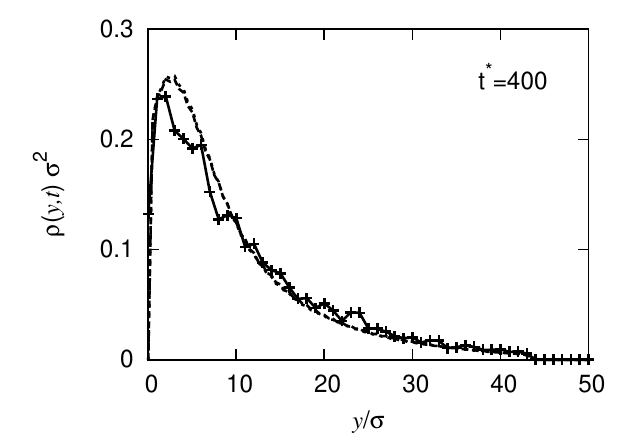}
\caption{\label{fig:E1_res}
Time evolution of the fluid density profiles $\rho(y,t)$. The solid lines are obtained from BD simulations, for $L=43.5\sigma$, $\beta a=0.1$, $\beta \epsilon = 1$ and there are $N=200$ particles in the system, corresponding to an average density $\bar{\rho}\sigma^2=0.072$. We also display results from DDFT (dashed line). Results are displayed at the times $t/\tau_B\equiv t^*=0$, 100 and 400, where $\tau_B=\beta \sigma^2/\Gamma$ is the Brownian time-scale.}
\end{figure*}

\begin{figure*}
\centering
\includegraphics[width=0.32\columnwidth]{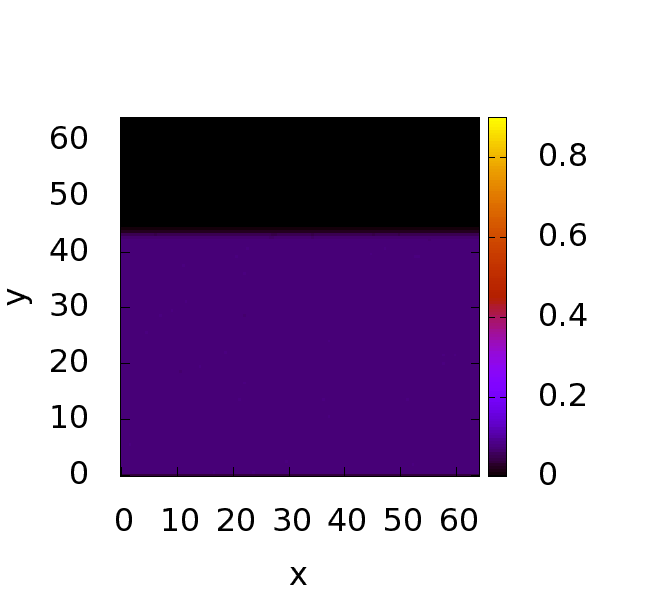}
\includegraphics[width=0.32\columnwidth]{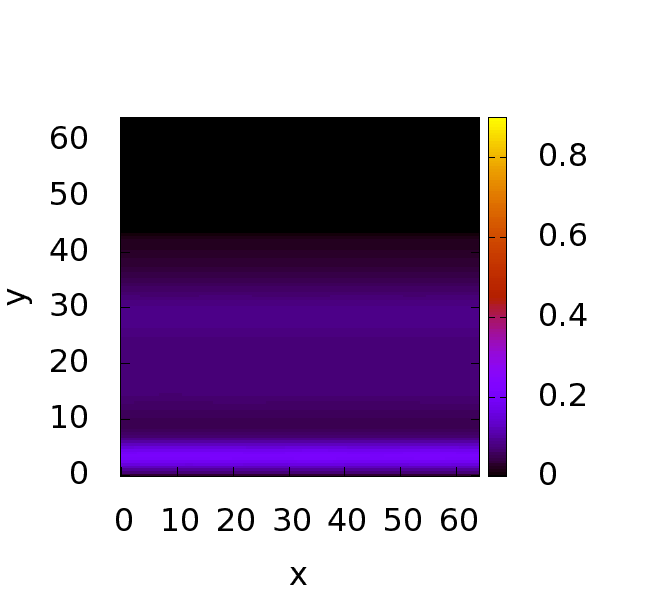}
\includegraphics[width=0.32\columnwidth]{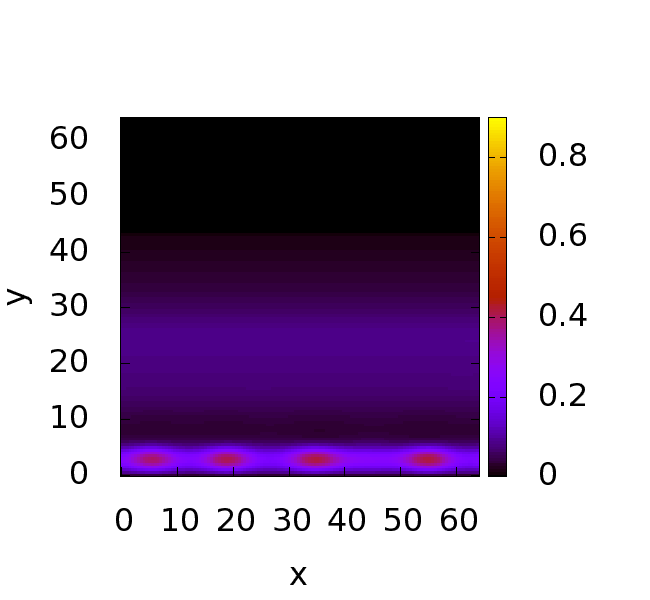}

\includegraphics[width=0.32\columnwidth]{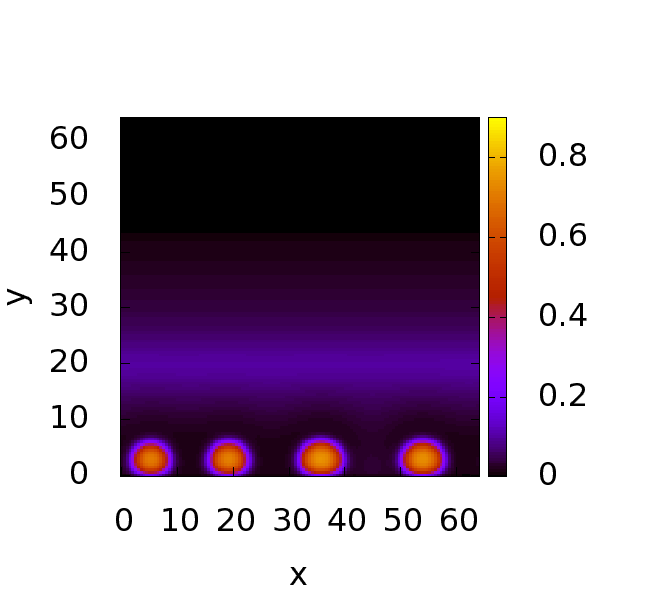}
\includegraphics[width=0.32\columnwidth]{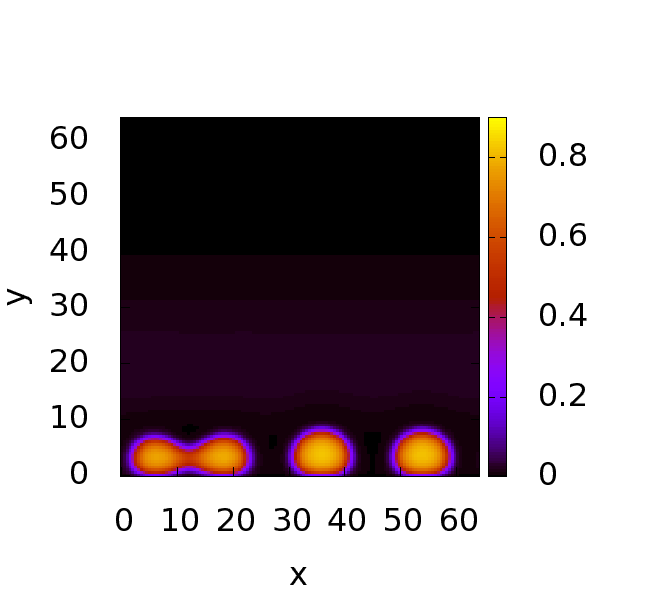}
\includegraphics[width=0.32\columnwidth]{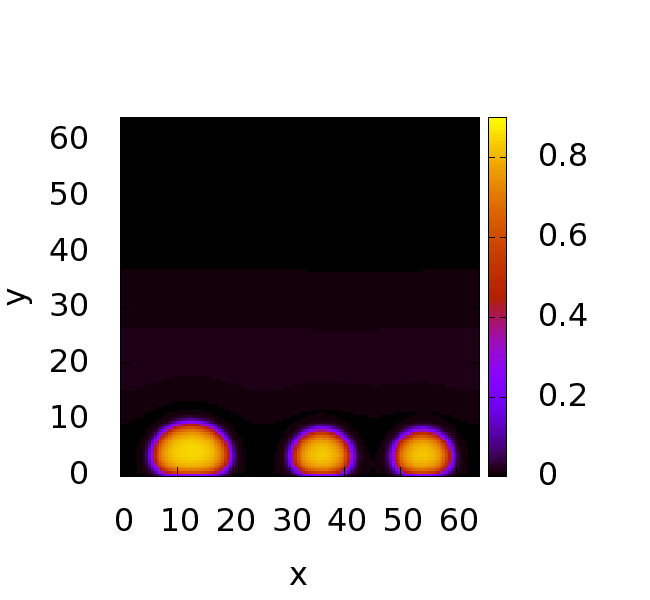}
\caption{\label{fig:DDFT_profiles_E3_5}
Fluid density profiles obtained from DDFT for the case when $L=43.5\sigma$, $\beta a=0.1$, $\beta \epsilon = 3.5$ and $N=200$ ($\bar{\rho}\sigma^2=0.072$) at the times $t^*=0$, 50, 75, 100, 250 and 300 (starting top left and finishing bottom right). These results should be compared with the BD simulation results displayed in Figs.\ \ref{fig:snapshots} and \ref{fig:BD_profiles}.}
\end{figure*}

\begin{figure*}
\centering
\includegraphics[width=0.32\columnwidth]{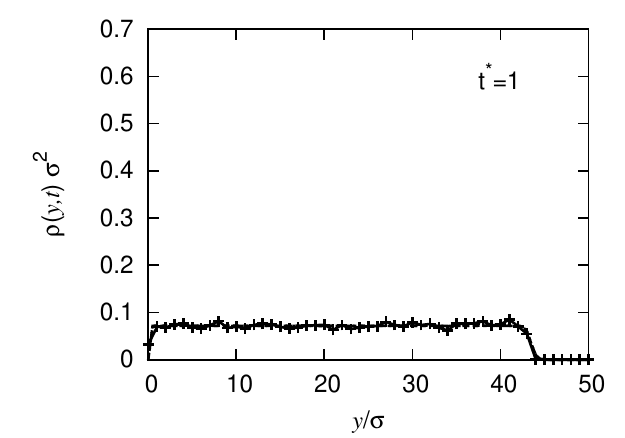}
\includegraphics[width=0.32\columnwidth]{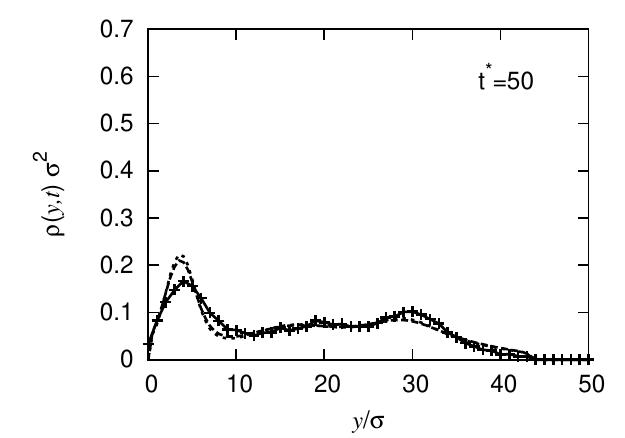}
\includegraphics[width=0.32\columnwidth]{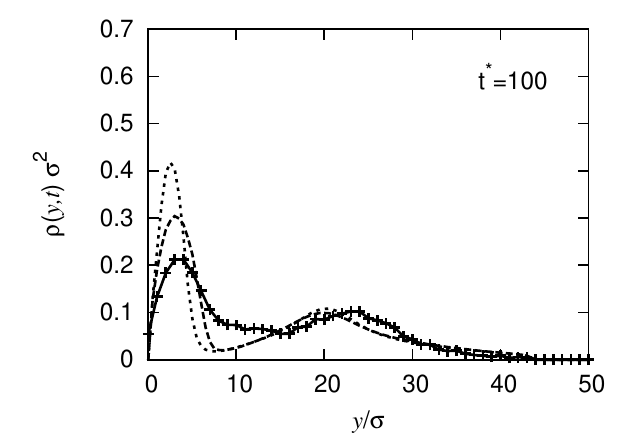}

\includegraphics[width=0.32\columnwidth]{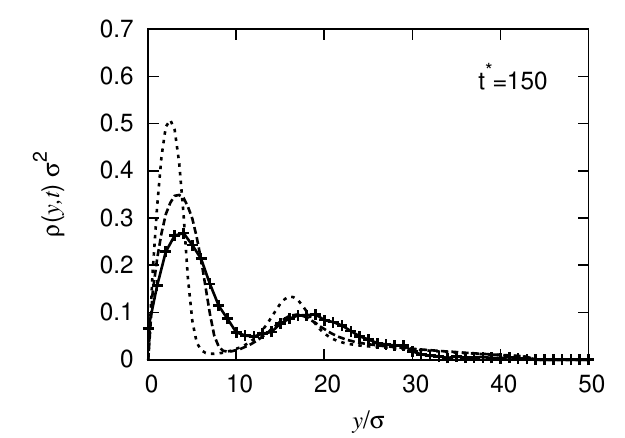}
\includegraphics[width=0.32\columnwidth]{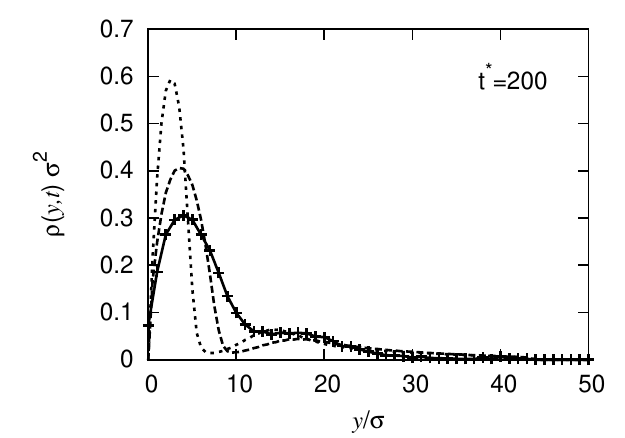}
\includegraphics[width=0.32\columnwidth]{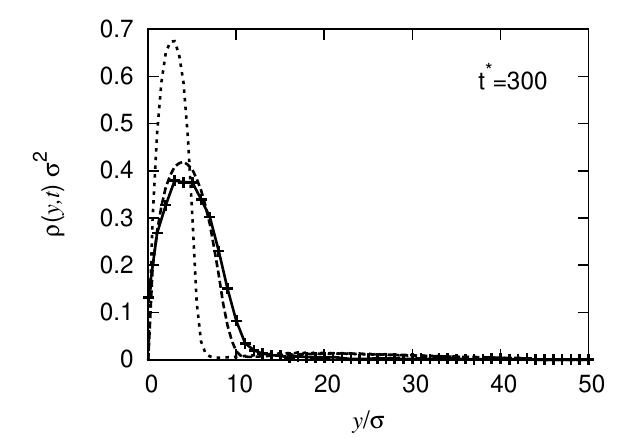}
\caption{\label{fig:DDFTvBD} Time evolution of the fluid density
profiles $\rho(y,t)$. The solid line are obtained from BD
simulations, for the same parameters as the results in Figs.\
\ref{fig:snapshots} and \ref{fig:BD_profiles}. We also display
results from a 1D DDFT calculation (dotted line) and also by
averaging over the $x$-direction the results from the 2D DDFT
calculation displayed in Fig.\ \ref{fig:DDFT_profiles_E3_5}.
Results are displayed at the times $t/\tau_B\equiv t^*=1$, 50,
100, 150, 200 and 300, where $\tau_B=\beta \sigma^2/\Gamma$ is the
Brownian time-scale.}
\end{figure*}

\begin{figure*}
\centering
\includegraphics[width=0.32\columnwidth]{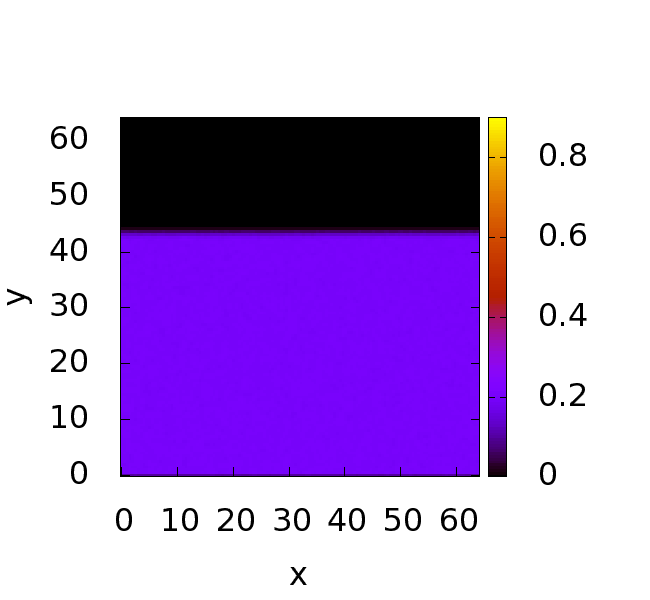}
\includegraphics[width=0.32\columnwidth]{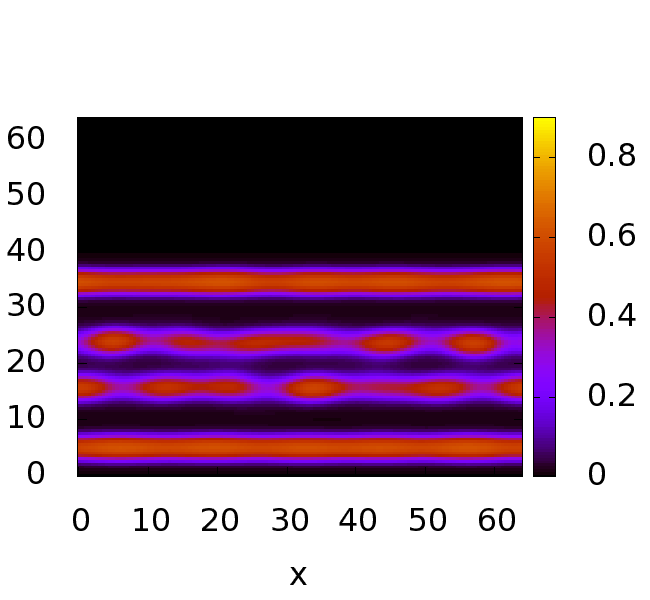}
\includegraphics[width=0.32\columnwidth]{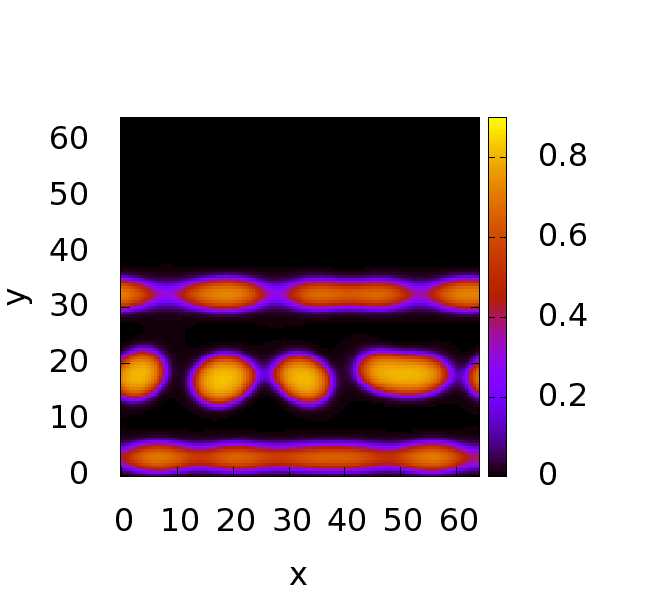}

\includegraphics[width=0.32\columnwidth]{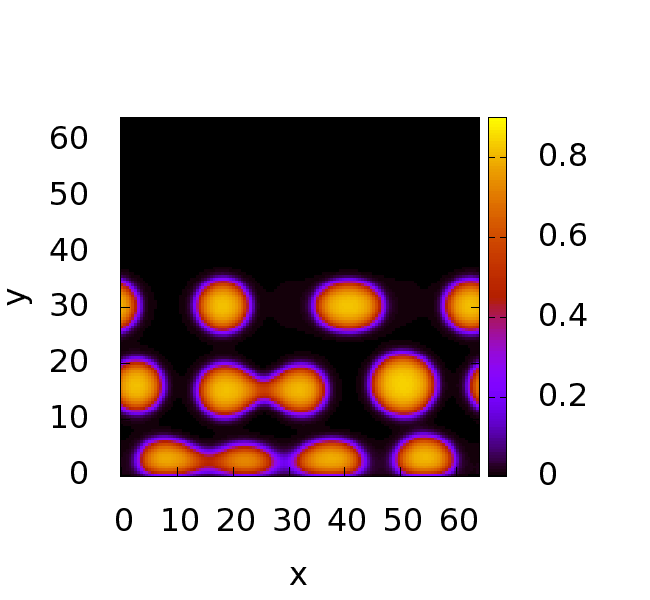}
\includegraphics[width=0.32\columnwidth]{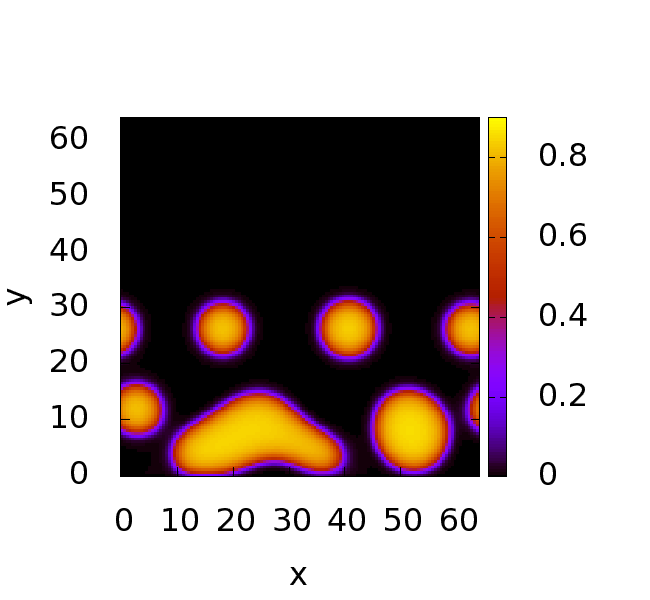}
\includegraphics[width=0.32\columnwidth]{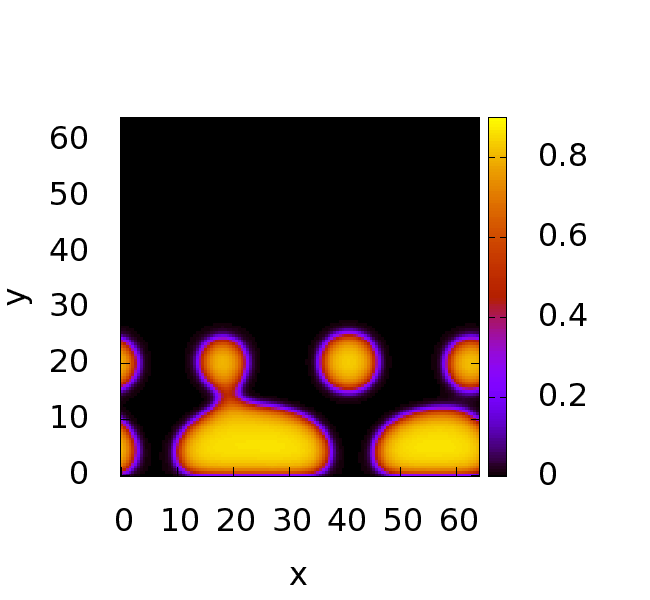}

\includegraphics[width=0.32\columnwidth]{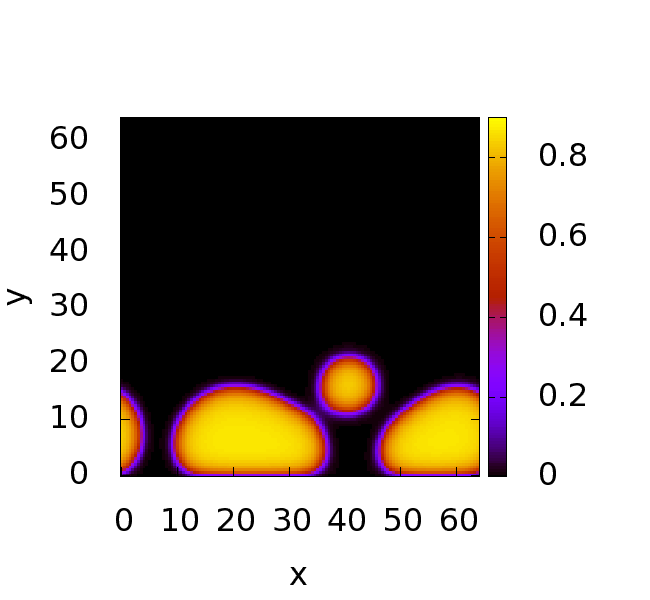}
\includegraphics[width=0.32\columnwidth]{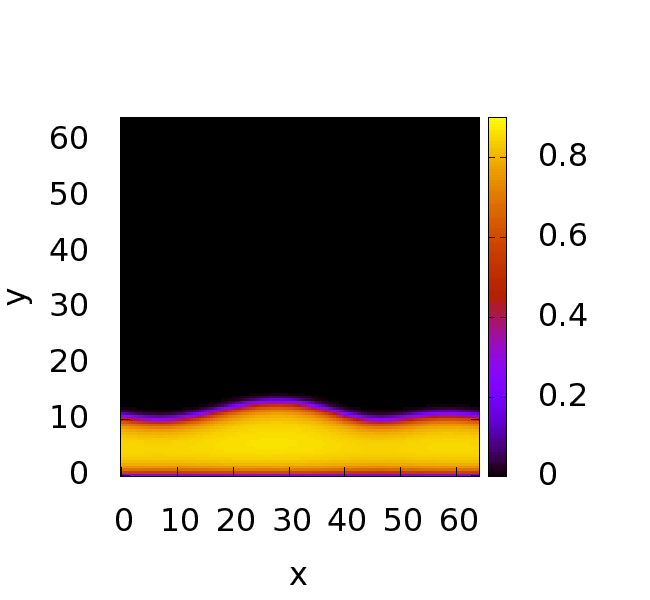}
\includegraphics[width=0.32\columnwidth]{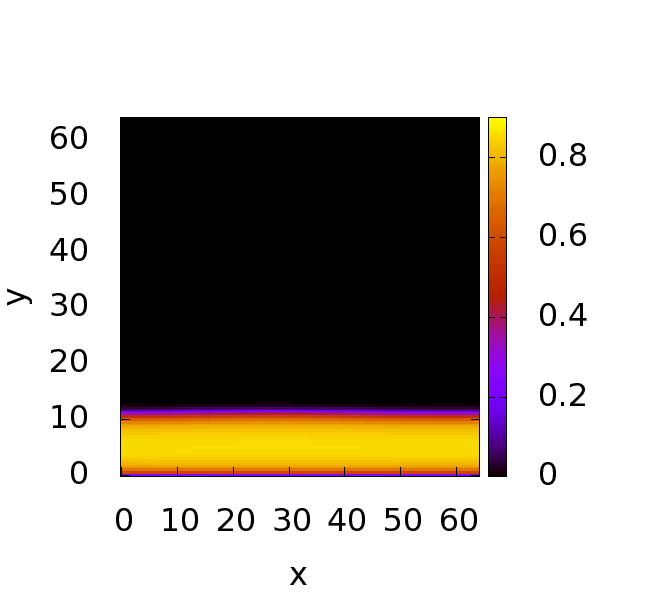}
\caption{\label{fig:DDFT_0_2}
Fluid density profiles obtained from DDFT for the case when $\beta a=0.1$, $\beta \epsilon = 3.5$ and $\bar{\rho}\sigma^2=0.2$ at the times $t^*=1$, 20, 40, 60, 100, 160, 200, 260 and $\infty$ (starting top left and finishing bottom right).}
\end{figure*}

We solve the DDFT by discretizing the equations on a cartesian
grid, with grid spacing $\Delta x=\Delta y=0.5\sigma$, and we set
the initial time $t=0$ density profile to be that of the ideal-gas
in the equilibrium situation when the drive in the external
potential $a=0$ (i.e.\ $\rho(x,y)=\bar{\rho}$, for $0<y<L$), which
has added to it a small random number at each grid point.
This random number is drawn from a
uniform distribution between $\pm\bar{\rho}/20$, where
$\bar{\rho}$ is the average fluid density in the system. The
results do not depend on the amplitude of this noise. The random
noise is added to the initial time $t=0$ density profile in order
to break the symmetry of the profile which subsequently in some
situations allows the system to develop density profiles that vary
in both the $x$- and in the $y$-directions, when the DDFT is
solved in 2D. Note that the density profiles from the DDFT do not
vary in the $x$-direction in all situations; they only vary in
this direction in certain specific situations -- see below.

One can also {\em assume} that the density profiles only vary in the
$y$-direction (which is strictly true, since the external
potential only varies in the $y$-direction) and use this fact to
reduce solving Eq.\ \eqref{eq:DDFT}, to that of solving for a
one-dimensional (1D) time series of density profiles $\rho(y,t)$.
Below, we compare results from doing this, to the results obtained
from the full 2D solution to the DDFT.

\section{Results}
\label{sec:results}

All the results displayed in this section are for the case when
the parameters in the external potential \eqref{eq:ext_pot} take
the values $L=43.5\sigma$ and $\beta a=0.1$. In Fig.\
\ref{fig:E1_res} we display results for the case when there are
200 particles in the system, which corresponds to an average
density $\bar{\rho}\sigma^2=0.072$, and when $\beta \epsilon=1$.
This level of attraction between the particles is not strong enough for the system to exhibit
liquid-gas phase separation as can be inferred from Fig.\
\ref{fig:phase_diag}. The system takes a time $t^* \sim 400$ to
reach equilibrium, where $t^* \equiv t/\tau_B$, and where
$\tau_B=\beta \sigma^2/\Gamma$ is the Brownian time scale, which
is the time it takes for a particle to diffuse a distance of order
its own diameter. Since the amplitude of the driving force $a$ is
rather small, we see that the density near the top of the system
$\rho\sigma^2\sim 0.01$ is not insignificant. This also comes as a
result of the attraction between the particles being quite
weak. We see that the agreement between the BD simulation results
and the DDFT results is fairly good. Note that in this situation,
the DDFT produces exactly the same result whether it is solved in
1D or in 2D. This is in contrast to the results which we display
below for cases where $\beta \epsilon$ is sufficiently large for
phase separation to be observed.

In Fig.\ \ref{fig:DDFT_profiles_E3_5} we display density profiles
obtained from the 2D DDFT for the case when $\beta \epsilon = 3.5$
and $\bar{\rho}\sigma^2=0.072$ at various times during the
sedimentation process. These results should be compared with the
BD simulation results (which are for $\beta \epsilon = 4.8$)
displayed in Figs.\ \ref{fig:snapshots} and \ref{fig:BD_profiles}.
We see that the theory predicts that the system prefers to form a
number of droplets, rather than form a layer of uniform thickness
along the bottom. Strictly speaking, since $\rho(\rr,t)$ is an
ensemble average quantity and since the external potential only
varies in the $y$-direction, these profiles should also only vary
in the $y$-direction. It is rather striking, however, that the
resulting density profiles in-fact resemble more closely what we
observe in the simulation snapshots in Fig.\ \ref{fig:snapshots}.

We also calculate the density profiles in this situation using the
1D DDFT. We display these results in Fig.\ \ref{fig:DDFTvBD},
together with the results from the 2D DDFT that have {\em
subsequently} been averaged over the $x$-direction to produce a 1D
density profile. Initially, for $t^*\lesssim 50$ there is no
difference between the results from these two approaches. However,
for latter times, when the 2D DDFT predicts that the system breaks
up into a number of separate drops, the results from the two
approaches are, of course, different. We also display in  Fig.\
\ref{fig:DDFTvBD} the BD simulation results (c.f.\ Figs.\
\ref{fig:snapshots} and \ref{fig:BD_profiles}), and remarkably the
results from the 2D DDFT that are subsequently averaged over
the $x$-direction are in much better agreement with the BD
simulation results then the 1D DDFT results. This suggests that
the additional degree of freedom in the 2D DDFT helps to improve the
theory, by `allowing' it to better describe the density
inhomogeneities that are present in the $x$-direction.

In Fig.\ \ref{fig:DDFT_0_2} we display density profiles obtained
from the 2D DDFT for the case when $\beta \epsilon = 3.5$ and
$\bar{\rho}\sigma^2=0.2$, at various times during the
sedimentation process. The average density in this case is higher
than in the case considered above in Fig.\
\ref{fig:DDFT_profiles_E3_5}. We see that the initially uniform
fluid breaks up into droplets of the liquid phase, due to the
attraction between the particles. Due to the external potential,
these droplets sediment to the bottom of the system and coalesce
to form a single uniform layer along the bottom of the system.
This is in contrast to the case in Fig.\
\ref{fig:DDFT_profiles_E3_5}, and is due to the fact that in the
case in Fig.\ \ref{fig:DDFT_0_2} there are more particles in the
system. This difference is best understood by considering the fact
that as the density of the fluid is increased, the area of the
surface that is covered by the two-dimensional liquid, which we denote $A$, also
increases. As discussed above in Sec.\ \ref{sec:BD_res}, the
interfacial tension between the liquid and the surface is roughly
equal to the interfacial tension $\gamma$ between the liquid and
the gas. Thus, as the number of particles (average density) in the
system is increased, the ratio $l/A$ decreases, where $l$ is the
length of the interface between the liquid and the gas plus the
length of the interface between the liquid and the wall (i.e.\ it
is the total length of the interface(s) around the 2D liquid). As the
system seeks to minimise the free energy, it therefore seeks to
minimise the contribution due to these interfaces $\sim \gamma l$.
Thus, the final equilibrium configuration is essentially
determined by minimising $l$ subject to the constraint that $A$ is
fixed. When the ratio $l/A$ is large (low $\bar{\rho}$), the system must form drops
to minimise $l$; this is the case in Fig.\
\ref{fig:DDFT_profiles_E3_5}. However, when $l/A$ is small (i.e.\ larger $\bar{\rho}$), $l$ is
minimised by the system forming a layer of uniform thickness along
the bottom; this is the case in Fig.\ \ref{fig:DDFT_0_2}. Note,
however, that this argument only applies in the limit when $a$ is
small. As the amplitude of the external potential $a$ is
increased, then the external potential energy contribution to the free
energy overcomes the interfacial tension contribution $l \gamma$
and the particles are squeezed to the bottom of the system, as can
be seen in Fig.\ \ref{fig:snapshots2}. This argument should also
apply to three-dimensional (3D) systems in a similar situation, as long as
one instead considers the ratio ${\cal A}/V$, where ${\cal A}$ is the surface area
of the liquid and $V$ is its volume. In this case, the system
will seek to minimise the free energy by balancing the interfacial
contribution~$\sim \gamma {\cal A}$, against the
external potential energy contribution.

\section{Concluding remarks}
\label{sec:conc}

We have shown that the behaviour during sedimentation of a 2D
model colloidal suspension with attraction between the particles
is rather rich due to the interplay between the aggregation stemming from
the particle attractions and the sedimentation due to the external
potential. We have used DDFT to describe the dynamics of the
system, and as the results in Figs.\ \ref{fig:E1_res} and
\ref{fig:DDFTvBD} show, the theory is in quite good agreement
with the simulation results. The agreement is particularly good
when the colloids are undergoing pure sedimentation (Fig.\
\ref{fig:E1_res}). However, even for the case displayed in Fig.\
\ref{fig:DDFTvBD}, where in addition to the sedimentation, the
particles are also aggregating due to the strong attractions
between them, we still observe good qualitative agreement between
the DDFT and the simulations.

The results from our work raise a key question: why does the DDFT
when it is solved in 2D, and then afterwards averaged over the
$x$-direction, give better agreement with the simulations than
when the DDFT is solved in 1D? As discussed above in Sec.\
\ref{sec:profile_nature}, solving equilibrium DFT is in principle
equivalent to evaluating the partition function for the system,
and therefore since our external potential varies in only the
$y$-direction, the density profile obtained by averaging over the
ensemble of all configurations of the system, must also only vary
in the $y$-direction. This should at least be the case for the
final equilibrium density profile obtained from the DDFT, since
recall that the equilibrium $t \to \infty$  solution of the DDFT is the solution
of Eq.\ \eqref{eq:F_min} -- i.e.\ it is a minimum of the free
energy \cite{archer2004ddf} and is equivalent to solving
equilibrium DFT. However, as can be seen in the final frame for
$t^*=300$ in Fig.\ \ref{fig:DDFTvBD}, this is not the case. The
final frame of Fig.\ \ref{fig:DDFTvBD} shows that the results from
the 2D calculation are in very good agreement with the simulation
results, whereas the results from the 1D calculation do not agree
with the simulation results.

When our DDFT (DFT) is solved in 2D, we believe the reason that
it does not always give a density profile that
only varies in the $y$-direction, is due to the fact that we are
not using the (unknown) exact free energy functional, but instead,
we are using an approximation for this quantity -- specifically
that given in Eq.\ \eqref{eq:F_our}. This approximate free energy
functional neglects some of the fluctuation contributions to the
free energy. This manifestation has been known for many years now,
in particular, in the context of using DFT to calculate the density profile for the free
interface between a liquid and its vapour. It is known
that DFT fails to include some of the interfacial fluctuations 
\cite{evans79}. See also the more recent work in Refs.\
\cite{ISI:000272311000089,  ISI:000260859300030,
ISI:000235394200040, ISI:000186068300035, ISI:000171637100029} and
references therein. This issue has also been discussed in a very
illuminating manner by Reguera and Reiss
\cite{RegueraReissJCP2004}. In Ref.\ \cite{RegueraReissJCP2004}
the behaviour of drops of a 3D liquid, with a fixed number of particles and
confined within a spherical cavity were studied
using both DFT and simulation. In terms of the physics involved,
the situation studied in Ref.\ \cite{RegueraReissJCP2004} is somewhat akin to the
situation we study here, but with the external potential parameter $a$ in
our model set to zero. Using an approximation for
the Helmholtz free energy functional that is the 3D
analogue of the functional in Eq.\ \eqref{eq:F_our}, the authors of
Ref.\ \cite{RegueraReissJCP2004} showed that
the DFT is unable to describe the fluctuations corresponding to drop
translations within the cavity and also fluctuations corresponding to
variations in the number of particles in a liquid drop.
By comparing their DFT results with Monte-Carlo simulation results for a system
where the centre of mass is constrained to coincide with the centre
of the cavity (i.e.\ so that the translational fluctuations are suppressed)
they found better agreement between the DFT and the simulation results.
In other words, the DFT describes well the density distribution within the drop(s)
but it is unable to describe the translational fluctuations of the
drop(s). This is also the conclusion that we draw on the basis of the results in Fig.\
\ref{fig:DDFTvBD} and leads us to surmise that when DFT is implemented
together with an approximate free energy functional, such as that in
Eq.\ \eqref{eq:F_our}, it leads to one obtaining an average over a subsection of phase space,
rather than averaging over all of phase space.

In Ref.\ \cite{RegueraReissJCP2004} the authors also discuss the relationship between DFT and field theory (FT). In FT, the central quantity is a Hamiltonian $E[\phi(\rr)]$, which is a functional of the order-parameter $\phi(\rr)$, which may be interpreted as being a coarse-grained density profile (See Ref.\ \cite{CiachPRE08} for a possible definition of this quantity). When solving the FT to obtain the order parameter profile $\phi(\rr)$ which corresponds to the {\em most probable} configuration of the system, one is required to solve the following Euler-Lagrange equation \cite{RegueraReissJCP2004}:
\begin{equation}
\frac{\delta E[\phi(\rr)]}{\delta \phi(\rr)} = \mu,
\label{eq:F_min_FT}
\end{equation}
[c.f.\ Eq.\ \eqref{eq:F_min}] where $\mu$ is the Lagrange-multiplier that originates from the constraint from fixing the total number of particles in the system, which mathematically plays the same role as the chemical potential. Now, of course, the functional $E[\phi(\rr)]$ is not known exactly and so one is required to make an approximation for this quantity based on physical insight and any other knowledge we have about the system of interest. In DFT, there is a similar process used to construct an approximation for the free energy functional $F[\rho]$ \cite{evans79, evans1992fif, Hansen2006tsl}. The end result of these two processes can be very similar or even the same; i.e.\ one considers $\rho(\rr)$ and $\phi(\rr)$ to be the same quantity (strictly speaking, they are not the same, and $\rho(\rr)=\langle \phi(\rr)\rangle$, i.e.\ $\rho(\rr)$ is a statistical average over all possible configurations of the coarse-grained density $\phi(\rr)$ \cite{archer2004ddf_b, CiachPRE08}) and therefore one is tempted to make the same approximations in constructing $E[\phi]$ as are made in constructing $F[\rho]$. See Ref.\ \cite{CiachPRE08} for an example of a FT constructed along these lines.

With these arguments in mind, it is now clear why our 2D DDFT results agree better with the simulation results than the 1D DDFT results: Our approximation for the free energy \eqref{eq:F_our} is unable to describe the translational fluctuations of drops in the system. However, by solving in 2D we obtain density profiles which correspond to particular likely configurations of the system. By subsequently averaging this density profile over the $x$-direction, we are averaging over several droplet profiles, and so we get better agreement with the simulations than the 1D DDFT because we are averaging over more `typical' configurations. In this respect, we are treating the DFT more like a FT. If we fully take the FT point of view, then the natural extension of our approach is to perform several different DDFT calculations for each situation, with each calculation having a different realisation of the random noise that we add to the initial density profile (see Sec.\ \ref{sec:DDFT}). One would then average over all of these different density profiles and presumably therefore average over more of phase space (i.e.\ averaging over more possible fluid configurations). We believe that doing this may give even better agreement with the simulation results than our present 2D DDFT results. We should add, however, that following this approach would be rather computationally expensive, particularly if one were investigating a 3D system -- it may be quicker to just simulate the system.

We believe that if we used a more accurate DFT such as Rosenfeld's fundamental measure theory \cite{rosenfeldPRA1990} in our 2D DDFT calculations, we would still see the same `symmetry breaking' (i.e.\ the theory would predict that the density profiles have variations in the $x$-direction). Using this alternative approximation for the free energy functional would undoubtably improve the description of the fluid structure by doing a better job of describing the short range correlations between the particles. However, what is really required for the problem here is a DFT that includes and averages over the large scale density fluctuations; in particular those corresponding to modes such as a droplet of particles translating along the bottom of the system. Constructing such a DFT requires developing a theory which is capable of describing fluctuations on several different length scales, which is a notoriously difficult problem in physics.

A final point that we should emphasise is that in both our simulations and in our theory we have not included the hydrodynamic interactions that play an important role in the dynamics of real colloids \cite{dhont1996idc}. It is possible to include these interactions between the particles within the DDFT approach \cite{RL2008prl, RL2009epe}. However, we leave this as future work.

\section*{Acknowledgements}

It is a pleasure to dedicate this paper to Professor Bob Evans, whose contributions to liquid state theory are immense. He has inspired several generations of scientists in the field and it has been a privilege and a joy to have worked with him. We wish him all the best in the years to come.

AJA gratefully acknowledges financial support from RCUK and AM acknowledges financial support of the MSMT of the Czech Republic under Project No.\ LC512 and the GAAS of the Czech Republic (Grant No. IAA400720710).

\bibliographystyle{apsrev}

\end{document}